\documentclass{tMOP2e}
\usepackage{amsmath}
\usepackage{amssymb}
\usepackage{graphicx}
\usepackage[english]{babel}

\begin{document}
\title{Impact of interparticle dipole-dipole interactions on optical nonlinearity of nanocomposites}
\author{Andrey V. Panov$^{a,b}$ $^{\ast}$\thanks{$^\ast$Corresponding author. Email: panov@iacp.dvo.ru}\\
\vspace{6pt}
$^{a}${\em{Institute of Automation and Control Processes,
Far Eastern Branch of Russian Academy of Sciences,
5, Radio st., Vladivostok, 690041, Russia}};\\
$^{b}${\em{School of Natural Sciences, Far Eastern Federal University, 8, Sukhanova st., Vladivostok, 690950, Russia}}\\
\vspace{6pt}
\received{Received: March 21, 2013}}
\jvol{60} \jnum{11} \jyear{2013} \jmonth{11}
\doi{10.1080/09500340.2013.822590} \issn{1362-3044} \issnp{0950-0340}
\setcounter{page}{915}
\maketitle
\begin{abstract}
In this paper, effect of dipole-dipole interactions on nonlinear optical properties of the system of randomly located semiconductor nanoparticles embedded in bulk dielectric matrix is investigated. This effect results from the nonzero variance of the net dipole field in an ensemble.
The analytical expressions describing the contribution of the dipole-dipole coupling to nonlinear dielectric susceptibility  are obtained. The derived relationships are applicable over the full range of nanoparticle volume fractions. The factors entering into the contribution and depending on configuration of the dipoles are calculated for several cases. It is shown that for the different arrangements of dipole alignments the relative change of this contribution does not exceed $1/3$.

This is an Author's Original Manuscript of an article whose final and definitive form, the Version of Record, has been published in the Journal of Modern Optics, Volume 60, Issue 11, Pages 915--919, 2013 12 Aug 2013 \copyright\ Taylor \& Francis, available online at: http://www.tandfonline.com/doi/abs/10.1080/09500340.2013.822590.
\end{abstract}


\section{Introduction}

Recently, nonlinear optical properties of nanocomposites have attracted much interest due to their potential applications in high-tech devices. In particular, these nanocomposites are promising substance for photovoltaics, lasers, optical switches and limiters. The nanocomposites exhibit the nonlinear characteristics much enhanced as against the homogeneous bulk material.
Typically, these nanocomposites comprise metallic or semiconductor nanoparticles embedded in a dielectric matrix \cite{Shalaev02,Beecroft97}. Up to now, the studies of the optical nonlinearity of the dielectric nanoparticles are quite rare \cite{Smith82,Lee09,Dzyuba11,Milichko13}.
An increase in the nonlinear optical properties of the metal-dielectric composites arises from localized surface plasmon resonances while the semiconductor nanocomposites enhance their optical nonlinearity due to quantum confinement effect.
Among possible sources of the composite optical nonlinearity, there can occur dipole-dipole interactions between polarized nanoparticles. The impact of the dipole-dipole interactions increases with rise in nanoparticle concentration in the composite. The effect of the dipole-dipole interactions on the optical characteristics of the nanocomposite is the most thoroughly studied in the case of the metallic inclusions in the matrix. As a rule, these investigations are accomplished by numeric methods \cite{Shalaev96,Myroshnychenko08}. One of the popular techniques, the discrete dipole approximation, also known as the coupled dipole approximation, can be exploited for simulating clusters of spheres \cite{Yurkin07}. Antosiewicz et al. \cite{Antosiewicz12} developed the similar approach based on replacing the retarded dipole sum by an integral. This procedure permitted them to describe analytically optical response of two-dimensional amorphous metallic nanoparticle arrays with interparticle coupling which were observed by experiment \cite{Antosiewicz12,Schwind12}. It is worth emphasizing that the coupled dipole method, being a linear approximation, is suitable for weak optical fields.

Little attention is given to study the effects of the interparticle dipole-dipole interactions in the semiconductor-dielectric nanocomposites. Normally, the semiconductor nanoparticles have sizes comparable to the exciton Bohr radius and possess energy level structure. External light radiation with some probability excites the nanoparticle in the state with dipole moment, the decay time of this state is much larger than the period of an optical field oscillation. A model describing the effect of the dipole-dipole interactions on the nonlinear optical properties of such nanocomposites was proposed in Ref.~\cite{Panov10}. This model deals with the system of semiconductor nanoparticles randomly arranged in the dielectric matrix. It is assumed that all the excited nanoparticles have equal absolute value $p$ of the electric dipole moment which is proportional to the incident light electric field amplitude $E_{in}$,
\begin{equation}
p = \varepsilon_m \alpha v E_{in},
\label{p:Ein}
\end{equation}
where $\alpha$ is the dimensionless polarizability of the particle, $v$ is its volume, $\varepsilon_m$ is the dielectric function of the matrix. The particle dipole moment $\mathbf{p}$ is assumed to be arbitrarily oriented in space. This continuous approximation can be justified for the nanoparticles with a number of exciton levels. The lifetime of the excited state is large in comparison with the period of an optical field, so that the dipoles can be treated in quasi-static approximation.  By this means, the $N$ excited nanocrystals will induce on the test particle, being located at the coordinate origin, additional random electric field
\[
\mathbf{E}=\sum_{l=1}^N\boldsymbol{\xi}_{l}\left( \mathbf{p}_l,\mathbf{r}_l \right),\]
where $\boldsymbol{\xi}_{l}\left( \mathbf{p}_l,\mathbf{r}_l \right)$ is the dipolar field from the $l$-th particle,
\begin{equation}
\boldsymbol{\xi}_{l}=\frac{3(\mathbf{r}_l\cdot\mathbf{p}_l)\mathbf{r}_l-\mathbf{p}_l r_l^2}{\varepsilon_m r_l^5}
\end{equation}
$\mathbf{r}_l$ is the position vector of the $l$-th particle and $\mathbf{p}_l$ is its dipole moment. As a first approximation, we assume that the net dipolar field $\mathbf{E}$ does not change the polarizability $\alpha$ of the individual particles. This field is random since the nanoparticles are polarized and placed arbitrarily. In Ref.~\cite{Panov10}, previously obtained distribution functions of the net dipolar field projection $E$ \cite{Grant64} was utilized. For volume concentrations (volume fractions) of the excited nanoparticles $c>0.1$ the Gaussian distribution was used as distribution functions of the random field $E$. 

In this work, we employ for bulk samples the negative cumulant expansion proposed for the two-dimensional distribution of dipoles \cite{surfdip,inter2d}. This will allow us to obtain the third-order self-induced optical susceptibility for volume fractions $c\lessapprox 0.1$ which frequently occur in experiments. Moreover, we will calculate the third-order susceptibilities for various angular configurations of the dipoles.

\section{Model}

\begin{figure}
{\centering\includegraphics[scale=1]{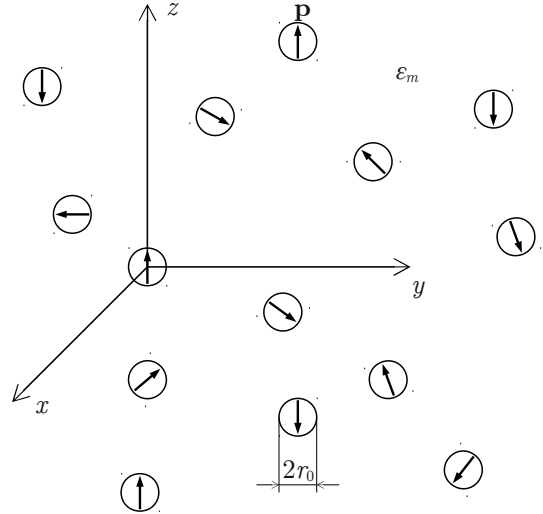}\par}
\caption{Schematic representation of the nanocomposite under consideration.\label{schem}}
\end{figure}
Let us consider an ensemble of identical spherical particles randomly arranged in the dielectric matrix which is illustrated in Fig.~\ref{schem}. The radius of the particles is $r_0$. The external laser radiation with some probability causes transition of the particles to excited state with dipole moment. After some period, a large portion of the particles will be in polarized state. In order to proceed, we should obtain distribution function $W(E)$ of the net dipolar random field projection $E$ onto the selected direction. The use of the field projection instead of vectors allows us to discuss without loss of generality several angular configurations of the dipoles. In a general way, this distribution function can be written as
\begin{equation}
W (E ) d E = \int \delta \left[E - \sum^N_{l =
1}{\xi}_l \right]\prod^N_{l =
1} \tau_l\left( \mathbf{p}_l \right)\, \mathrm{d} \mathbf{p}_{l}\, \mathrm{d}\mathbf{r}\, \mathrm{d} E ,
\end{equation}
where $\tau_l \left( \mathbf{p}_l \right)$ is the distribution function for dipole moment orientations, $\delta$ is the Dirac $\delta$-function. We assume that due to macroscopic isotropy $\tau_l$ is identical for all the particles, $\tau_l \left( \mathbf{p}_l \right)=\tau\left( \mathbf{p}\right)$, $\mathbf{p}_l=\mathbf{p}$. After applying Markov's method \cite{Chandra43}, we obtain a characteristic function 
\[A(\rho)=\int\limits_{-\infty}^\infty W(E) \exp(i\rho E) \mathrm{d}E=\exp\left[-C(\rho) \right] \]
  with
\begin{equation}
 C(\rho)=\frac{c}{v}\int\left[ 1-\exp\left(i\rho {\xi}\right) \right] \tau\left( \mathbf{p}\right)
 \mathrm{d} \mathbf{p}\, \mathrm{d}\mathbf{r},
 \label{initint}
\end{equation}
where $\rho$ is the Fourier-transform variable corresponding to the field projection $E$, the volume concentration of excited nanoparticles $c=N v/V$, $V$ is the volume of macroscopic sample, $N$ is the number of the particles with the dipole moment. This approach is employed for theoretical treatment of the dipole-dipole interactions in the resonance methods \cite{Grant64}, dilute magnetic systems \cite{Klein68} and ferroelectric dipole glasses \cite{Vugmeister90}. In a fashion similar to the two-dimensional model \cite{surfdip}, we express $C(\rho)$ in powers of $\rho$ (the negative cumulant expansion). Thus, the distribution function $W(E)$ after the inverse Fourier transform can be formulated in integral form,
\begin{equation}
W(E) =\frac{1}{2\pi}\int\limits_{-\infty}^\infty\exp\biggl\lbrace-\frac{c}{v}\biggl[ \frac{\lambda_2}{2}\rho^2-\frac{\lambda_4}{4!}\rho^4 
+\frac{\lambda_{6}}{6!}\rho^{6}+\ldots\biggr]-i\rho E  \biggr\rbrace \mathrm{d}\rho,
\label{wseriesbulk}
\end{equation}
where the cumulants $\lambda_n$ are defined as follows
\begin{equation}
\lambda_n=\frac1{p^n}\int \left\lbrace \frac{3(\mathbf{r}\cdot\mathbf{p})\mathbf{r}-\mathbf{p}r^2}{r^5}\right\rbrace ^n \tau(\mathbf{p})\mathrm{d} \mathbf{p}\, \mathrm{d}\mathbf{r}.
\label{defmun}
\end{equation}
For instance, $\lambda_1$ and $\lambda_2$ define the mean field and the variance of the dipolar field projection, $\lambda_4$ determines its fourth moment and so forth. As stated before,
the absolute value of $\mathbf{p}$ is equal for all the particles, but they have different angular components of $\mathbf{p}$. In spherical coordinates $\mathbf{r}=(r,\theta,\phi)$, integration over radius in Eq.~(\ref{defmun}) begins from $2r_0$ which is the minimum distance between two sphere centers. The cumulant expansion in Eq.~(\ref{wseriesbulk}) is done under the assumption that the number of the dipoles with one direction is balanced by dipoles with opposite alignment owing to the random nature of the excitement. Thus, this expansion has only even-numbered terms \cite{surfdip}, the odd-numbered $\lambda_n$ are zero.

Then, following Ref.~\cite{Panov10}, after Gibbs averaging, we obtain the density of free energy of the ensemble
\begin{equation}
\label{freeener}
 F=-k_B T  \frac{c}{v}\ln \int\limits_{-\infty}^\infty \exp\left\lbrace -\frac{p E}{k_B T}\right\rbrace W(E) \mathrm{d}E,
\end{equation}
where $T$ is the temperature and $k_B$ is the Boltzmann constant, and the projection of macroscopic polarization $P$ onto the selected direction, resulting from the dipole-dipole interactions,
\begin{equation}
 P=-\frac{\partial \langle F\rangle_t}{\partial E_{in}},
 \label{therm_relP}
\end{equation} 
where $\langle \rangle_t$ denotes time averaging.  For linear or circular light polarization, by expressing $P$ as a series in powers of the applied optical field amplitude $E_{in}$, we get terms containing linear and nonlinear optical susceptibilities. On doing so, under the assumption $\varepsilon_m>0$, one can derive
\begin{multline}
P=\frac{c \alpha^2 v \varepsilon_m^2}{ k_B T I_{0,0}^2}
\Biggl\lbrace \left[\frac{c}{v}\lambda_2 (k_B T)^2 I_{2,0}-I_{0,2}\right]I_{0,0}+\\I_{0,1}^2\Biggr\rbrace E_{in}+
\frac{c \alpha^3 v^2 \varepsilon_m^3}{2 (k_B T)^2 I_{0,0}^3}\Biggl\{\left[ 3 \frac{c}{v} (k_B T)^2I_{1,2}\lambda_2-{I_{0,3}}\right]I_{0,0}^2 - \\
3\left[ \frac{c}{v}(k_B T)^2I_{0,2}\lambda_2-{I_{0,2}}\right]I_{0,1}I_{0,0} -2{I_{0,1}^3} \Biggr\}E_{in}^2+\\
\frac{c^2 \alpha^4 v^3 \varepsilon_m^4}{6 k_B TI_{0,0}^4} 
\Biggl\{-
I_{0,0}^3\biggl[\left(\lambda_4+3\lambda_2^2\frac{c}{v}\right)(k_B T)^2 I_{4,0}-\\ 6I_{2,2}\lambda_2+\frac{v}{c(k_B T)^2}I_{0,4}\biggr]+ 
4 I_{0,0}^2 I_{0,1} \left[\frac{vI_{0,2}}{c(k_B T)^2} - 3I_{2,0}\lambda_2\right]+ \\
12 I_{0,0} I_{0,1}^2  \left[I_{2,0}\lambda_2 - \frac{vI_{0,2}}{c(k_B T)^2}\right]+\\
3 I_{0,0}^2 \frac{c}{v}\left[I_{2,0}\lambda_2 k_B T - \frac{vI_{0,2}}{c k_B T}\right]^2 + 6 I_{0,1}^4 
\Biggr\}E_{in}^3+\ldots,
\label{Pseries}
\end{multline}
where
\begin{equation}
I_{k,l}=\int_{-\infty}^\infty \int_{-\infty}^\infty \exp(-i\rho E)\rho^k E^l \mathrm{d}\rho \mathrm{d} E.
\label{Erhoint}
\end{equation}
Upon employing the Fourier integral representation of the Dirac $\delta$-function and the definition of the Dirac $\delta$-function derivative of the $n$-th order, one can reveal that integrals (\ref{Erhoint})
have nonzero values only if $k=l$ and $k=0,2,4,\ldots$, namely $I_{0,0}=-2\pi$, $I_{2,2}=4\pi$ \cite{inter2d}.

After simplifications the first non-zero term in (\ref{Pseries}) contains $E_{in}^3$ which gives us the formula for the self-induced Kerr nonlinear susceptibility:
\begin{equation}
\chi^{(3)} = -\frac{2 c^2 \alpha^4 v^2 \varepsilon_m^4 \lambda_2}{k_B T}.
\label{chi3-3d}
\end{equation}
It should be noticed that the same relationship for $\chi^{(3)}$ can be obtained using the Gaussian distribution as $W(E)$, however, this result has not been obvious initially since the employment of the Gaussian distribution is justified for large concentrations of the dipoles. It is worth emphasizing that Eq.~(\ref{chi3-3d}) can be applied not only to the dipole-dipole interactions but also to other types of interparticle couplings. In the latter case,  $\lambda_2$ will be defined by other formulas. Since the variance $\lambda_2>0$, the contribution of interparticle interactions into the self-induced Kerr nonlinear susceptibility is always negative.

If one continue expansion in (\ref{Pseries}) then the next nonzero term will contain $\chi^{(7)}$:
\begin{equation}
\chi^{(7)} = -\frac{ c^2 \alpha^8 v^6 \varepsilon_m^8 \lambda_4}{3(k_B T)^3}.
\label{chi7-2d}
\end{equation}

In the statement of this model, we have inferred that the dipole moment of the particles linearly depends on the incident field amplitude (\ref{p:Ein}). We may abandon this approximation replacing Eq.~(\ref{p:Ein}) with
\begin{equation}
p = \varepsilon_m \alpha v E_{in} + \chi^{(3)}_p E_{in}^3,
\label{p:Ein3}
\end{equation}
where $\chi^{(3)}_p$ is the third-order hyperpolarizability of the individual nanoparticle. After repeating all the calculations, it can be shown that $\chi^{(3)}_p$ makes a contribution to the fifth-order nonlinearity of the system $\chi^{(5)}$.

\section{Angular configurations of dipoles}

In the following, we discuss several dipole arrangements. When all the excited dipoles are collinear with two equiprobable opposite orientations,
\begin{equation}
\tau(\gamma,\psi)=\frac{\delta(\gamma)+\delta(\gamma-\pi)}{4\pi},
\label{tauIsing3D}
\end{equation}
where $\gamma$ and $\psi$ are the polar and azimuthal angles specifying $\mathbf{p}$. Such a configuration of the dipoles is expected when the linearly polarized light wave propagates through the nanocomposite. Then 
\begin{multline}
\lambda_2 = \int\limits_{2r_0}^\infty r^2 \mathrm{d}r\int\limits_0^{\pi}\sin\theta \mathrm{d}\theta\int\limits_0^{2\pi}\mathrm{d}\phi\int\limits_0^{\pi}\sin\gamma \mathrm{d}\gamma\int\limits_0^{2\pi}\mathrm{d}\psi \:\tau(\gamma,\psi)\times\\ \left\{\frac{3[\sin\gamma\cos(\phi-\psi)\sin\theta+\cos\gamma\cos\theta]\cos\theta-\cos\gamma}{r^3\varepsilon_m}\right\}^2.
\end{multline}
After integrating,
\begin{equation}
\lambda_2 = \frac{ 16 \pi }{15 \left( 2 r_0 \right)^3 \varepsilon_m^2}
\label{mu2-3disingpar}
\end{equation}
and on substituting into Eq.~(\ref{chi3-3d})
\begin{equation}
 \chi^{(3)} = -\frac{4\pi\varepsilon_m^2 v^2 \alpha^4 c^2}{15 r_0^3 k_B T}.
 \label{chi3gau}
\end{equation}
This formula was derived in Ref.~\cite{Panov10} under the assumption of Gaussian distribution as $W(E)$. 

In the case of arbitrarily oriented dipoles
\begin{equation}
\tau(\gamma,\psi)=\frac 1{4\pi}
\end{equation} 
and integration in (\ref{defmun}) gives
\begin{equation}
\lambda_2 = \frac{ 8 \pi }{9 \left( 2 r_0 \right)^3 \varepsilon_m^2}.
\label{mu2-3darbitrary}
\end{equation}

Further, let us consider the dipoles chaotically oriented  with $\mathbf{p}$ lying in the $x$-$y$ plane, i.e. $\mathbf{p}$ has only $x$ and $y$ components. This arrangement can occur when the nanocomposite is embedded in planar waveguide with propagating TM-polarized wave. In this case, we select the direction in the $x$-$y$ plane, e.g. $x$. Then, the distribution function for dipole moment orientations becomes
\begin{equation}
\tau(\gamma,\psi)=\frac {\delta(\gamma-\pi/2)}{2\pi}
\end{equation} 
and
\begin{equation}
\lambda_2 = \frac{ 4 \pi }{5 \left( 2 r_0 \right)^3 \varepsilon_m^2}.
\label{mu2-3darbitraryx-y}
\end{equation}

It should be underlined that the relative change of $\lambda_2$ in (\ref{mu2-3disingpar}), (\ref{mu2-3darbitrary}), and (\ref{mu2-3darbitraryx-y}) does not exceed $33$\% despite the high anisotropy of the dipolar field. This fact can be attributed to the random placement of the nanoparticles in the matrix. Thus, the variation of the dipole arrangements affect insignificantly the contribution of the dipole-dipole interactions to the third-order nonlinear optical susceptibility of the nanocomposite.

\section{Discussion}

Typically, the II-VI semiconductor quantum dots in the excited state possess the large values of the electric dipole moment \cite{Colvin92,WangF06}. Hence, it is reasonable to make a comparison to the nanocomposites containing such quantum dots. Given $\alpha=0.4$ \cite{WangF06}, $c=0.02$, $2r_0=4$~nm, $v=4\pi r_0^3/3$, $\varepsilon_m=1.8$, $T=300$~K, using Eqs.~(\ref{chi3-3d}), (\ref{mu2-3darbitrary}) one can calculate that $\chi^{(3)} \approx -8\times 10^{-11}$~esu.
This value is of the same order as observed by experiment with nanosecond laser
pulses for cadmium chalcogenide nanocomposites
($-10^{-11}\ldots-10^{-10}$~esu) \cite{Jing09,Du02}. As a rule, under femtosecond
laser pulses, the nanocomposites with cadmium chalcogenide quantum dots show the positive real part of the
third-order susceptibility \cite{Gan08,Jin09,Zhu11}, but at nanosecond pulse durations
they display the negative sign of that \cite{Jing09,Du02}. This
phenomenon may be attributed to the influence of the interparticle dipole-dipole coupling on the third-order susceptibility as this effect will become substantial upon exciting enough quantity of the nanoparticles. At shorter timescales other fast optical nonlinearities (e.g. two-photon absorption, free-carrier absorption and dispersion, bound electronic Kerr effect, quantum-confined Stark effect
) will prevail.

It should be stressed that the susceptibility nonlinearity (\ref{chi3-3d}) in this model arises from the random field dispersion that yields nonzero $\lambda_2$ and higher order moments in Eq.~(\ref{wseriesbulk}). Namely, $\lambda_2$ makes a contribution to the third-order susceptibility, $\lambda_4$ does to $\chi^{(7)}$ and so forth. This result cannot be obtained within the framework of mean-field theories including effective-medium theories which lack the field dispersion. In other words, the mean-field models has only one distribution function moment $\lambda_1$ (the mean field). Such theories can only translate the nonlinearity of one constituent to the effective nonlinearity of the mixture \cite{Stroud88}.

Notice, that in contrast to Ref.~\cite{Panov10} in this work the expressions for $\chi^{(3)}$ are obtained without restriction on certain ranges of nanoparticle volume fractions. The expression for $\chi^{(3)}$ calculated in Ref.~\cite{Panov10} for the case of $c\rightarrow 0$ is almost twice than $\chi^{(3)}$ computed
with Eq.~(\ref{chi3gau}) but the first was obtained under the rather rough limitation of the maximum field value in the use of the Cauchy-Lorentzian distribution.

\section{Conclusions}

In summary, the analytical expressions for the contribution of the dipole-dipole coupling $\chi^{(3)}$ to the Kerr optical nonlinearity of the system of the randomly located nanoparticles have been
derived for the nanoparticle volume concentrations of several percent. This contribution is caused by the dispersion of the random dipolar field. The dipolar induced contribution to the third-order susceptibility is calculated for different types of dipole configurations which are connected with the light
polarization. It has been shown that for the different dipole arrangements $\chi^{(3)}$ varies moderately. 


\begin{thebibliography}{26}
\providecommand{\natexlab}[1]{#1}

\bibitem[1]{Shalaev02}
Shalaev, V.M. (Ed.),   Optical properties of nanostructured random media. In:
  {\em Topics in Applied Physics};  Vol. 82,   Springer: Berlin Heidelberg,
  2002.

\bibitem[2]{Beecroft97}
Beecroft, L.L.; Ober, C.K. Nanocomposite materials for optical applications.
  {\em Chem. Mater.}  {\bf 1997}, {\em 9} (6), 1302--1317.

\bibitem[3]{Smith82}
Smith, P.W.; Maloney, P.J.; Ashkin, A. Use of a liquid suspension of dielectric
  spheres as an artificial {Kerr} medium.  {\em Opt. Lett.}  {\bf 1982}, {\em
  7} (8) (Aug), 347--349.

\bibitem[4]{Lee09}
Lee, W.M.; {El-Ganainy}, R.; Christodoulides, D.N.; Dholakia, K.; et~al.
  Nonlinear optical response of colloidal suspensions.  {\em Opt. Express}
  {\bf 2009}, {\em 17} (12) (Jun), 10277--10289.

\bibitem[5]{Dzyuba11}
Dzyuba, V.; Milichko, V.; Kulchin, Y. Nontypical photoinduced optical
  nonlinearity of dielectric nanostructures.  {\em J. Nanophoton.}  {\bf 2011},
  {\em 5} (1), 053528.

\bibitem[6]{Milichko13}
Milichko, V.A.; Dzyuba, V.P.; Kulchin, Y.N. Unusual nonlinear optical
  properties of {SiO$_2$} nanocomposite in weak optical fields.  {\em Appl.
  Phys. A}  {\bf 2013}, {\em 111} (1), 319--322.

\bibitem[7]{Shalaev96}
Shalaev, V.M. Electromagnetic properties of small-particle composites.  {\em
  Phys. Rep.}  {\bf 1996}, {\em 272}, 61--137.

\bibitem[8]{Myroshnychenko08}
Myroshnychenko, V.; Rodriguez-Fernandez, J.; Pastoriza-Santos, I.; Funston,
  A.M.; Novo, C.; Mulvaney, P.; Liz-Marzan, L.M.; et~al. Modelling the optical
  response of gold nanoparticles.  {\em Chem. Soc. Rev.}  {\bf 2008}, {\em 37}
  (9), 1792--1805.

\bibitem[9]{Yurkin07}
Yurkin, M.; Hoekstra, A. The discrete dipole approximation: an overview and
  recent developments.  {\em Journal of Quantitative Spectroscopy and Radiative
  Transfer}  {\bf 2007}, {\em 106} (1--3), 558--589.

\bibitem[10]{Antosiewicz12}
Antosiewicz, T.J.; Apell, S.P.; Z\"ach, M.; Zori\ifmmode~\acute{c}\else
  \'{c}\fi{}, I.; et~al. Oscillatory optical response of an amorphous
  two-dimensional array of gold nanoparticles.  {\em Phys. Rev. Lett.}  {\bf
  2012}, {\em 109} (Dec), 247401.

\bibitem[11]{Schwind12}
Schwind, M.; Miljkovi\'c., V.D.; Z\"ach, M.; Gusak, V.; K\"all, M.; Zori\'c,
  I.; et~al. Diffraction from arrays of plasmonic nanoparticles with
  short-range lateral order.  {\em ACS Nano}  {\bf 2012}, {\em 6} (11),
  9455--9465.

\bibitem[12]{Panov10}
Panov, A.V. Contribution of dipolar interactions to third-order nonlinear
  dielectric susceptibility of nanocomposites.  {\em Opt. Lett.}  {\bf 2010},
  {\em 35} (11), 1831--1833.

\bibitem[13]{Grant64}
Grant, W.J.C.; Strandberg, M.W.P. Statistical theory of spin-spin interactions
  in solids.  {\em Phys. Rev.}  {\bf 1964}, {\em 135} (3A), A715--A726.

\bibitem[14]{surfdip}
Panov, A. Probability distribution function of dipolar field in two-dimensional
  spin ensemble.  {\em Eur. Phys. J. B}  {\bf 2012}, {\em 85} (4), 122.

\bibitem[15]{inter2d}
Panov, A.V. Effect of dipolar interactions on optical nonlinearity of
  two-dimensional nanocomposites.  {\em J. Opt.}  {\bf 2013}, {\em 15} (5)
  (May), 055201.

\bibitem[16]{Chandra43}
Chandrasekhar, S. Stochastic problems in physics and astronomy.  {\em Rev. Mod.
  Phys.}  {\bf 1943}, {\em 15} (1) (Jan), 1--89.

\bibitem[17]{Klein68}
Klein, M.W. Temperature-dependent internal field distribution and magnetic
  susceptibility of a dilute {Ising} spin system.  {\em Phys. Rev.}  {\bf
  1968}, {\em 173} (2), 552--561.

\bibitem[18]{Vugmeister90}
Vugmeister, B.E.; Glinchuk, M.D. Dipole glass and ferroelectricity in
  random-site electric dipole systems.  {\em Rev. Mod. Phys.}  {\bf 1990}, {\em
  62}, 993--1026.

\bibitem[19]{Colvin92}
Colvin, V.L.; Alivisatos, A.P. {CdSe} nanocrystals with a dipole moment in the
  first excited state.  {\em J. Chem. Phys.}  {\bf 1992}, {\em 97} (1),
  730--733.

\bibitem[20]{WangF06}
Wang, F.; Shan, J.; Islam, M.; Herman, I.; Bonn, M.; et~al. Exciton
  polarizability in semiconductor nanocrystals.  {\em Nature Materials}  {\bf
  2006}, {\em 5} (11), 861--864.

\bibitem[21]{Jing09}
Jing, C.; Xu, X.; Zhang, X.; Liu, Z.; et~al. In situ synthesis and third-order
  nonlinear optical properties of {CdS/PVP} nanocomposite films.  {\em J. Phys.
  D: Appl. Phys.}  {\bf 2009}, {\em 42} (7), 075402.

\bibitem[22]{Du02}
Du, H.; Xu, G.Q.; Chin, W.S.; Huang, L.; et~al. Synthesis, characterization,
  and nonlinear optical properties of hybridized {CdS} polystyrene
  nanocomposites.  {\em Chem. Mater.}  {\bf 2002}, {\em 14} (10), 4473--4479.

\bibitem[23]{Gan08}
Gan, C.; Zhang, Y.; Liu, S.; Wang, Y.; et~al. Linear and nonlinear optical
  refractions of {CR39} composite with CdSe nanocrystals.  {\em Optical
  Materials}  {\bf 2008}, {\em 30} (9), 1440--1445.

\bibitem[24]{Jin09}
Jin, Q.; Wu, W.; Zheng, Z.; Yan, Y.; Liu, W.; Li, A.; Yang, Y.; et~al. The
  third-order optical nonlinearity and upconversion luminescence of {CdTe}
  quantum dots under femtosecond laser excitation.  {\em J. Nanopart. Res.}
  {\bf 2009}, {\em 11} (3), 665--670.

\bibitem[25]{Zhu11}
Zhu, B.H.; Zhang, H.C.; Zhang, J.Y.; Cui, Y.P.; et~al. Surface-related
  two-photon absorption and refraction of {CdSe} quantum dots.  {\em Appl.
  Phys. Lett.}  {\bf 2011}, {\em 99} (2), {021908}.

\bibitem[26]{Stroud88}
Stroud, D.; Hui, P.M. Nonlinear susceptibilities of granular matter.  {\em
  Phys. Rev. B}  {\bf 1988}, {\em 37} (15) (May), 8719--8724.

\end{thebibliography}

\end{document}